\documentclass[twoside,12pt]{report}
\usepackage{graphicx}
\usepackage{amssymb, amsmath, amsxtra}
\usepackage{epsfig,subfigure}

\usepackage{latexsym}
\usepackage{bm}

 \usepackage{ntheorem}
           \theoremstyle{plain}
                      {\theorembodyfont{\rmfamily}
                      \theoremseparator{.}

           \theoremstyle{plain}
           
           \theoremstyle{plain} 
           \theoremstyle{plain}

           \theoremstyle{plain}
           
            }

\begin{document}
\begin{center}
{\Large \textbf{GROUP ANALYSIS OF NONLINEAR}\\[1ex] \textbf{INTERNAL WAVES IN
OCEANS}}\\[1.5ex]
 {\Large \textbf{III: Additional conservation laws}}\footnote{Published
 in \textit{Archives of ALGA}, vol. 6, 2009, pp. 55-62.}\\[1ex]
 {\sc Nail H. Ibragimov}\\
 Department of Mathematics and Science, Blekinge Institute
 of Technology,\\ 371 79 Karlskrona, Sweden\\[.5ex]
 {\sc Ranis N. Ibragimov}\\
Department of Mathematics, Research and Support Center for Applied
Mathematical Modeling (RSCAMM),\\
 New Mexico Institute of Mining and Technology,\\
 Socorro, NM, 87801 USA
 \end{center}

 \noindent
\textbf{Abstract.} Using the maximal Lie algebra of point symmetries
 of a system of nonlinear equations used in geophysical fluid dynamics,
 two conservation laws are found in addition to the
 conservation of energy.\\[2ex]


  \section{Introduction}
  \label{Acl:int}
 \setcounter{equation}{0}

 The maximal group of Lie point symmetries of the system
 \begin{align}
 \Delta \psi_t - g \rho_x - f v_z & =
 \psi_x \Delta \psi_z - \psi_z \Delta \psi_x\,, \label{SphNR.eq1}\\[1ex]
 v_t + f \psi_z & = \psi_x v_z - \psi_z v_x\,, \label{SphNR.eq2}\\[1ex]
 \rho_t + \frac{N^2}{g}\, \psi_x  & = \psi_x \rho_z - \psi_z \rho_x \label{SphNR.eq3}
 \end{align}
 has been presented in \cite{iik09}. It is generated by the infinite-dimensional
 Lie algebra spanned by the following operators:
 \begin{align}
  &
 X_1= \frac{\partial}{\partial v}  \,, \quad
 X_2= \frac{\partial}{\partial \rho}\,, \quad
 X_3= a(t) \frac{\partial}{\partial \psi}\,, \quad X_4= \frac{\partial}{\partial t} \,,\notag \\
  &
  X_5= b(t) \left[\frac{\partial}{\partial x} - f\,\frac{\partial}{\partial v}\right] +
       b'(t) z\,\frac{\partial }{\partial \psi}\,, \notag\\
       &
 X_6= c(t) \left[\frac{\partial }{\partial  z} + \frac{N^2}{g}\,
  \frac{\partial}{\partial \rho}\right] -
      c'(t) x\,\frac{\partial}{\partial \psi}\,,  \label{SphNR.eq4}\\
  &
 X_7= x \frac{\partial}{\partial x}  +
      z \frac{\partial}{\partial z} +
      v \frac{\partial}{\partial v} +
      \rho \frac{\partial}{\partial \rho}  +
      2 \psi \frac{\partial}{\partial \psi}\,, \notag\\ &
 X_8= t \frac{\partial }{\partial  t}  +
      2 x \frac{\partial }{\partial  x}  +
      2 z \frac{\partial }{\partial  z}  +
      3 \psi \frac{\partial }{\partial  \psi} -
      2 f x \frac{\partial}{\partial v} +
      2 \frac{N^2}{g}\, z \frac{\partial}{\partial \rho} \,, \notag\\ &
 X_9= z \frac{\partial}{\partial x} -
      x \frac{\partial}{\partial z} -
      \frac{1}{f}\,\big[g \rho + (f^2 - N^2) z\big]
      \frac{\partial}{\partial v}
     + \frac{1}{g}\,\big[fv + (f^2 - N^2) x \big]
       \frac{\partial}{\partial \rho}\,\cdot\notag
\end{align}

 \section{Conservation law provided by semi-dilation}
 \label{sd:cons}
 \setcounter{equation}{0}

 Consider the operator $X_8$ from (\ref{SphNR.eq4}),
 \begin{equation}
\label{SphNR.eq5}
  X_8= t \frac{\partial }{\partial  t}  +
      2 x \frac{\partial }{\partial  x}  +
      2 z \frac{\partial }{\partial  z}  +
      3 \psi \frac{\partial }{\partial  \psi} -
      2 f x \frac{\partial}{\partial v} +
      2 \frac{N^2}{g}\, z \frac{\partial}{\partial \rho}
 \end{equation}
 It generates the following one-parameter transformation group with the parameter
  $\varepsilon:$
 \begin{equation}
 \label{SphNR.eq6}
 \begin{split}
 & \bar t = t {\rm e}^\varepsilon, \quad \bar x = x {\rm e}^{2 \varepsilon}, \quad
 \bar z = z {\rm e}^{2 \varepsilon}, \quad \bar \psi = \psi {\rm e}^{3 \varepsilon},\\[1ex]
 & \bar v = v + f x \left(1 - {\rm e}^{2 \varepsilon}\right),
 \quad \bar \rho = \rho - \frac{N^2}{g}\left(1 - {\rm e}^{2 \varepsilon}\right).\\[1ex]
 \end{split}
 \end{equation}
 Since some of variables, namely $t, x, z$ and $\psi$ are
 subjected to
 dilations while two other variables transform otherwise,
 we call (\ref{SphNR.eq6}) the \textit{semi-dilation group}.
 Let us construct the conserved vector provided by this group.

 \subsection{Computation of the conservation law density}

 We will use the following formula for computing the density of
 conservation laws (see Eq. (3.23) in \cite{ibr-ibr09b})
 \begin{equation}
 \label{SphNR.eq7}
  C^1 = - v\, W^1 - \frac{g^2}{N^2}\,\rho\, W^2 - \psi_x\,D_x\big(W^3\big) - \psi_z\,D_z\big(W^3\big),
 \end{equation}
 where (see Eq. (3.16) in \cite{ibr-ibr09b})
 \begin{equation}
 \label{SphNR.eq8}
  W^\alpha = \eta^\alpha - \xi^j u_j^\alpha, \quad
  \alpha = 1, 2, 3.
  \end{equation}
  These formulas are written using the notation $t, \ x, \ z, \ v, \ \rho, \ \psi.$

 In the case of the operator (\ref{SphNR.eq5}) the quantities
(\ref{SphNR.eq8}) are written:
 \begin{equation}
 \label{SphNR.eq9}
 \begin{split}
 & W^1 = - 2 f x - t v_t - 2 x v_x - 2 z v_z,\\[1ex]
 & W^2 =  2 \frac{N^2}{g}\,z - t \rho_t - 2 x \rho_x - 2 z \rho_z,\\[1ex]
 & W^3 =  3 \psi - t \psi_t - 2 x \psi_x - 2 z \psi_z.\\[1ex]
 \end{split}
 \end{equation}
 Substituting (\ref{SphNR.eq9}) in (\ref{SphNR.eq8})
 we obtain upon simple calculations:
 \begin{align}
 \label{SphNR.eq10}
 &C^1 =  2 (fx v - g z\rho) - |\nabla \psi|^2 + \big(x D_x + z D_z\big)\Big(v^2
 + \frac{g^2}{N^2}\,\rho^2\Big)\\[1.5ex]
 &  + \big(x D_x + z D_z\big)\big(|\nabla \psi|^2\big)
 + t \left[v v_t + \frac{g^2}{N^2}\, \rho \rho_t + \psi_x \psi_{xt}
 + \psi_z \psi_{zt}\right].\notag
 \end{align}
 We can drop the last term in (\ref{SphNR.eq10}) because it
 can be written in the divergent form upon elimination of $v_t, \ \rho_4$ and
 $\psi_t$ by using Eqs. (\ref{SphNR.eq1})-(\ref{SphNR.eq3}).
 Indeed, it is shown in \cite{ibr-ibr09b}, Section 4.6, that the
 expression in the square brackets (cf. Eq. (4.11) in \cite{ibr-ibr09b})
 evaluated on the solutions of Eqs. (\ref{SphNR.eq1})-(\ref{SphNR.eq3})
 has the divergent form. Multiplication by $t$ does not violate this
 property. Then we use the identities
 \begin{align}
 \big(x D_x & + z D_z\big)\Big(v^2 + \frac{g^2}{N^2}\,\rho^2\Big)
 = - 2 \Big(v^2 + \frac{g^2}{N^2}\,\rho^2\Big)\notag\\[1ex]
 & \qquad \qquad \quad \ \ + D_x \Big[x\Big(v^2 + \frac{g^2}{N^2}\,\rho^2\Big)\Big]
 + D_z \Big[z\Big(v^2 + \frac{g^2}{N^2}\,\rho^2\Big)\Big], \notag\\[2ex]
 \big(x D_x & + z D_z\big)\big(|\nabla \psi|^2\big)
 = - 2 \psi|^2 + D_x \big[x\big(|\nabla \psi|^2\big)\big]
 + D_z \big[z\big(|\nabla \psi|^2\big)\big], \notag
  \end{align}
  drop the divergent type terms and
  obtain the following conserved density:
 \begin{equation}
 \label{SphNR.eq11}
 C^1 = 2 \left(fx v - g z\rho -\frac{1}{2}|\nabla \psi|^2 \right)
 - 2 \Big(v^2 + \frac{g^2}{N^2}\,\rho^2 + |\nabla \psi|^2\Big).
 \end{equation}

 Finally we note that the last term in (\ref{SphNR.eq11}) is the energy density
 (see Eq. (4.17) in \cite{ibr-ibr09b}). Therefore we eliminate it and conclude that
 the invariance under the semi-dilation with the generator (\ref{SphNR.eq5})
  provides the conservation law with the density
 \begin{equation}
 \label{SphNR.eq12}
 P = fx v - g z\rho -\frac{1}{2}|\nabla \psi|^2.
 \end{equation}

 \subsection{Conserved vector}

 Let us find the components $C^2, \ C^3$ of the conserved vector
 with the density (\ref{SphNR.eq12}). We will apply the procedure
 used in \cite{ibr-ibr09b}, Section 4.7. We have:
 $$
 D_t (P) = fx v_t - g z\rho_t -(\psi_x \psi_{xt} + \psi_z \psi_{zt}).
 $$
 Using Eqs. (\ref{SphNR.eq1})-(\ref{SphNR.eq3}), we obtain:
 \begin{align}
 & D_t (P)\Big|_{(\ref{SphNR.eq1})-(\ref{SphNR.eq3})} = - f^2 x \psi_z
 + fx \psi_x v_z - fx \psi_z v_x + N^2 z \psi_x\notag\\
 &- g z\psi_x \rho_z + g z\psi_z \rho_x
 - D_x(\psi \psi_{xt}) - D_z(\psi \psi_{zt}) + \psi \Delta \psi_t.\notag
 \end{align}
 One can rewrite this equation, using Eq. (4.23) from \cite{ibr-ibr09b},
in the following form:
 \begin{align}
  D_t (P)\Big|_{(\ref{SphNR.eq1})-(\ref{SphNR.eq3})} & =
 D_x\big(N^2 z \psi + fx \psi v_z - g z\psi \rho_z - \psi \psi_{xt}
 + \frac{1}{2} \psi^2 \Delta \psi_z\big)\notag\\[1ex]
& - D_z\big(f^2 x \psi + fx \psi v_x - g z\psi \rho_x + \psi
\psi_{zt} + \frac{1}{2} \psi^2 \Delta \psi_x\big).\notag
 \end{align}

Thus, the generator (\ref{SphNR.eq5})
  provides the conservation law
 $$
 D_t(P) + D_x(C^2) + D_z(C^3= 0
 $$
  with the density $P$ given by (\ref{SphNR.eq12}) and the
 flux given by the equations
 \begin{align}
 & C^2 = - N^2 z \psi - fx \psi v_z + g z\psi \rho_z + \psi \psi_{xt}
 - \frac{1}{2} \psi^2 \Delta \psi_z,\notag\\
 & C^3 = f^2 x \psi + fx \psi v_x - g z\psi \rho_x + \psi
\psi_{zt} + \frac{1}{2} \psi^2 \Delta \psi_x.\notag
 \end{align}

 \subsection{Conserved density $\bm{P}$ of the generalized invariant solution}

 If we substitute in (\ref{SphNR.eq12}) the generalized
 invariant solution (5.28)-(5.30) from our paper \cite{ibr-ibr09b},
 \begin{align}
 & \psi = A(\lambda) \cos (\omega t) + B(\lambda) \sin (\omega t),
 \notag\\[1ex]
 & v = \frac{fm}{\omega} \big[B'(\lambda) \cos (\omega t) -
 A'(\lambda) \sin (\omega t)\big], \notag\\[1ex]
 & \rho = \frac{k N^2}{g \omega} \big[B'(\lambda) \cos (\omega t) -
 A'(\lambda) \sin (\omega t)\big], \notag
 \end{align}
 we obtain:
 \begin{align}
 P = \frac{1}{\omega}\,(f^2 & mx  - N^2 k z) \big[B'(\lambda) \cos (\omega t) -
 A'(\lambda) \sin (\omega t)\big]\notag\\[1ex]
 & - \frac{k^2 + m^2}{2}
 \big[A'(\lambda) \cos (\omega t) + B'(\lambda) \sin (\omega
 t)\big]^2.\notag
 \end{align}

 \section{Conservation law provided by the rotation}
 \label{rot:cons}
 \setcounter{equation}{0}

 Taking the rotation generator $X_9$ from (\ref{SphNR.eq4})
 and proceedings as in Section \ref{sd:cons} we obtain the following conserved
 density:
 \begin{equation}
 \label{SphNR.rot1}
 Q = v \rho + f x \rho - \frac{N^2}{g}\, z v.
 \end{equation}

 Writing Eqs. (\ref{SphNR.eq1})-(\ref{SphNR.eq3}) by using the
 Jacobians $J(\psi, v) = \psi_x v_z - \psi_z v_x,$ etc.,  we have:
 \begin{align}
 \label{SphNR.rot2}
  D_t (Q)\Big|_{(\ref{SphNR.eq1})-(\ref{SphNR.eq3})}&  =
 v \left[J(\psi, \rho) - \frac{N^2}{g}\,\psi_x\right]
 + \rho \left[J(\psi, v) - f \psi_z\right]\\[1.5ex]
 & + fx \left[J(\psi, \rho) - \frac{N^2}{g}\,\psi_x\right]
 - \frac{N^2}{g}\, z \left[J(\psi, v) - f \psi_z\right].\notag
 \end{align}
The reckoning shows that
 \begin{align}
 & v J(\psi, \rho) + \rho J(\psi, v) = D_z(v \rho \psi_x) - D_x(v \rho \psi_z),\notag\\[1ex]
 & x J(\psi, \rho) - \rho \psi_z = D_z(x \rho \psi_x) - D_x(x \rho \psi_z),\notag\\[1ex]
 & z J(\psi, v) + v \psi_x = D_x(z v \psi_z) - D_z(z v \psi_x),\notag\\[1ex]
 & z \psi_z - x \psi_x = D_z(z \psi) - D_x(x \psi).\notag
 \end{align}
 Substituting these expressions in Eq. (\ref{SphNR.rot2}) we
 conclude that the rotation generator $X_9$ provides the conservation law
 $$
 D_t(Q) + D_x(C^2) + D_z(C^3= 0
 $$
  with the density $P$ given by (\ref{SphNR.rot1}) and the
 flux given by the equations
 \begin{align}
 & C^2 = \left[v \rho  + f x \rho  -
  \frac{N^2}{g}\, z v\right] \psi_z + \frac{N^2}{g}\,f z
  \psi,\notag\\[1.5ex]
 & C^3 = \left[\frac{N^2}{g}\, z v - v \rho - f x \rho\right] \psi_x - \frac{N^2}{g}\,f x \psi.\notag
 \end{align}

\newpage

 \vspace*{1cm}

 \section{Summary of conservation laws}
 \label{Acl:sum}
 \setcounter{equation}{0}

 It has been demonstrated in \cite{ibr-ibr09b} that the
 system of nonlinear equations (\ref{SphNR.eq1})-(\ref{SphNR.eq3})
 is self-adjoint. This property of the system has been used for
 deriving local conservation laws applying the method developed in
  \cite{ibr07a} to the infinitesimal symmetries (\ref{SphNR.eq4}).
  Some of the conservation laws associated with these symmetries
  are \textit{trivial}, i.e. have vanishing densities. But five
  conservation laws are nontrivial.

 The nontrivial conservation laws obtained in \cite{ibr-ibr09b} and in the present
  paper are summarized below. For the convenience of the reader, we
  formulate them both in the integral and differential
  forms.

 \subsection{Conservation laws in integral form}
 \label{Acl:clint}

 \begin{equation}
 \label{Acl:i1}
 \frac{d}{d t} \int\!\int v\, dx d z  = 0.\\[.5ex]
 \end{equation}

 \begin{equation}
 \label{Acl:i2}
 \frac{d}{d t} \int\!\int \rho\, dx d z  = 0.\\[.5ex]
 \end{equation}

 \begin{equation}
 \label{Acl:i3}
 \frac{d}{d t} \int\!\int \left[v^2 + \frac{g^2}{N^2}\, \rho^2
 + |\nabla \psi|^2\right] dx d z  = 0.\\[1.5ex]
 \end{equation}

 \begin{equation}
 \label{Acl:i4}
 \frac{d}{d t} \int\!\int \left[fx v - g z\rho -\frac{1}{2}|\nabla \psi|^2\right] dx d z
   = 0.\\[1.5ex]
 \end{equation}

 \begin{equation}
 \label{Acl:i5}
 \frac{d}{d t} \int\!\int \left[v \rho + f x \rho - \frac{N^2}{g}\, z v\right] dx d z
   = 0.\\[1.5ex]
 \end{equation}

 \subsection{Conservation laws in differential form}
 \label{Acl:cldif}

 \begin{equation}
 \tag*{(\ref{Acl:i1}$'$)}
 D_t(v) + D_x(v \psi_z) +D_z(f \psi - v \psi_x) = 0.
 \end{equation}

 \begin{equation}
 \tag*{(\ref{Acl:i2}$'$)}
 D_t(\rho) + D_x\left(\frac{N^2}{g}\, \psi + \rho\, \psi_z\right) +D_z(- \rho\, \psi_x) = 0.
 \end{equation}

 \begin{align}
  & D_t\left(v^2 + \frac{g^2}{N^2}\, \rho^2
 + |\nabla \psi|^2\right)\notag\\[1ex]
 & + D_x\left(2 g \rho \psi + v^2 \psi_z + \frac{g^2}{N^2}\,\rho^2 \psi_z
  - 2 \psi \psi_{xt} + \psi^2 \Delta \psi_z\right)\tag*{(\ref{Acl:i3}$'$)}\\[1ex]
 & + D_z\left(2 f v \psi - v^2 \psi_x - \frac{g^2}{N^2}\,\rho^2 \psi_x
  - 2 \psi \psi_{zt} - \psi^2 \Delta \psi_x\right) =
  0.\notag
 \end{align}

 \begin{align}
  & D_t\left(fx v - g z\rho -\frac{1}{2}|\nabla \psi|^2\right)\notag\\[1ex]
 & + D_x\left(- N^2 z \psi - fx \psi v_z + g z\psi \rho_z + \psi \psi_{xt}
 - \frac{1}{2} \psi^2 \Delta \psi_z\right)\tag*{(\ref{Acl:i4}$'$)}\\[1ex]
 & + D_z\left(f^2 x \psi + fx \psi v_x - g z\psi \rho_x + \psi
 \psi_{zt} + \frac{1}{2} \psi^2 \Delta \psi_x\right) =
 0.\notag
 \end{align}

 \begin{align}
  & D_t\left(v \rho + f x \rho - \frac{N^2}{g}\, z v\right)\notag\\[1ex]
 & + D_x\left(\Big[v \rho  + f x \rho  -
  \frac{N^2}{g}\, z v\Big] \psi_z + \frac{N^2}{g}\,f z
  \psi\right)\tag*{(\ref{Acl:i5}$'$)}\\[1ex]
 & + D_z\left(\Big[\frac{N^2}{g}\, z v - v \rho - f x \rho\Big] \psi_x
  - \frac{N^2}{g}\,f x \psi\right) = 0.\notag
 \end{align}

 The conservation law (\ref{Acl:i3}) defines the energy of the system.
 It seems that the conservation laws (\ref{Acl:i4})
 and (\ref{Acl:i5}), unlike (\ref{Acl:i3}), do not have direct analogies in mechanics
 and should be investigated from point of view of their physical significance. \\[4ex]
 \null \hfill 22 April 2009

 \end{document}